\documentclass[a4paper,draft]{amsproc}
\usepackage{amssymb}
\usepackage{amscd} 

\theoremstyle{plain}
 \newtheorem{thm}{Theorem}[section]
 \newtheorem{prop}{Proposition}[section]
 \newtheorem{lem}{Lemma}[section]
 \newtheorem{cor}{Corollary}[section]
\theoremstyle{definition}

\theoremstyle{remark}
 
 \numberwithin{equation}{section}

\title[Group classification of reaction-diffusion equations]{Group classification of variable coefficient quasilinear reaction-diffusion  equations}



\author[Vaneeva]{\bfseries Olena Vaneeva}

\address{
Department of Applied Research \\ 
Institute of Mathematics of NAS of Ukraine   \\ 
3~Tereshchenkivska Str., Kiev 01601\\
Ukraine}
\email{vaneeva@imath.kiev.ua}

\author[Zhalij]{Alexander Zhalij}
\address{Department of Applied Research \\ 
Institute of Mathematics of NAS of Ukraine   \\ 
3~Tereshchenkivska Str., Kiev 01601\\
Ukraine}
\email{zhaliy@imath.kiev.ua}

\begin{document}\allowdisplaybreaks

\vspace{18mm}
\setcounter{page}{1}
\thispagestyle{empty}

\begin{abstract}
The group classification of variable coefficient quasilinear reaction-diffusion equations $u_t=u_{xx}+h(x)B(u)$ is carried out exhaustively.
This became possible due to usage of a conditional equivalence group found in the course of the study of admissible point transformation within the class.
\end{abstract}

\maketitle

\section{Introduction}  

Geometrical study of differential equations (DEs) has a long and distinguished history dating back to the second part of XIX century when  the pioneering works of Gaston Darboux, Sophus Lie and \'Elie Cartan were published. Their ideas became a source for a number of developments including the theory of completely integrable systems and study of conservation laws. A brief but very nice review of both classical and modern treatments of geometrical study of differential equations is presented in~\cite{vaneeva:Kamran}.
One of the classical topics of such studies is Lie (point) symmetries of differential equations.
Lie proved that knowledge of continuous group of nondegenerate  point transformations that leave an equation invariant allows one to
reduce this equation to one with fewer independent variables. In many cases, this reduction procedure results in construction of a group-invariant solution in closed form. It is worthy to say that Lie symmetries represent by themselves powerful tool for finding exact solutions for partial differential equations (PDEs), and this is one of the most successful applications of geometrical studies of DEs~\cite{vaneeva:Olver1986,vaneeva:Ovsiannikov1982}.
Another  feature of Lie symmetries is that they reveal equations which are important for applications among wide set of admissible ones.
Indeed, all basic equations of mathematical physics, e.g., the equations
of Newton, Laplace, Euler--Lagrange, d'Alembert, Lam\'{e}, Hamilton--Jacobi, Maxwell, Schr\"odinger etc., have
rich symmetry properties~\cite{vaneeva:FN}. This property distinguishes these equations from other  PDEs. Therefore,  an important problem arises to single out from a given class of PDEs those
admitting Lie symmetry algebra of the maximally possible dimension. This problem is called \emph{ group
classification problem} and is formulated as follows~\cite{vaneeva:Ovsiannikov1982,vaneeva:Ibragimov1994V1}: given a class
of PDEs, to classify all possible cases of extension of Lie invariance  algebras of such
equations  with respect to the equivalence
group of the class.

At this stage the modern group analysis provides us with two main approaches for solving group classification problems. The first is algebraic one, based on
subgroup analysis of the corresponding equivalence group~\cite{vaneeva:Basarab-Horwath&Lahno&Zhdanov2001,vaneeva:Bihlo&Cardoso-Bihlo&Popovych2012}. It results in complete group classification only if the class under study is normalized~\cite{vaneeva:Popovych&Kunzinger&Eshraghi2010}. Roughly speaking, the class is normalized if any point transformation between two fixed equations from the class is induced by transformation from its equivalence group.
The second approach is based on direct integration of determining equations implied by the infinitesimal invariance criterion. It is efficient usually only for classes of simple structure having one or two arbitrary elements of a single variable. Obviously, normalized classes or ones of simple structure do not exhaust
a set of classes of PDEs or their systems which are important for application. To solve more group classification problems, a number of notions was introduced recently as well as new approaches were developed. These are, e.g., notions of admissible \cite{vaneeva:Popovych&Kunzinger&Eshraghi2010} (synonym:  form-preserving~\cite{vaneeva:Kingston&Sophocleous1998}) transformations, generalized~\cite{vaneeva:Meleshko1994} and extended~\cite{vaneeva:mogran} equivalence groups, normalized class of DEs~\cite{vaneeva:Popovych&Kunzinger&Eshraghi2010}, equivalence groupoid~\cite{vaneeva:Popovych&Bihlo2012}, contractions of equations and conservation laws~\cite{vaneeva:monstr2,vaneeva:VPS2012}. Among new approaches it is worthy to mention the method of furcate split~\cite{vaneeva:Popovych&Ivanova2004NVCDCEs}, the method of mapping between classes~\cite{vaneeva:VPS2009}, the partition of a class into normalized subclasses~\cite{vaneeva:Bihlo&Cardoso-Bihlo&Popovych2012,vaneeva:Popovych&Kunzinger&Eshraghi2010}, etc.

In this paper we solve the group classification problem for the class of variable coefficient semilinear reaction-diffusion equations
of the form
\begin{equation}\label{eqRDk}
u_t=u_{xx}+h(x)B(u),
\end{equation}
where $h=h(x)$ and $B=B(u)$ are arbitrary smooth functions of their variables,
$hB_{uu}\neq0$. Linear equations singled out of class~\eqref{eqRDk} by the condition $B_{uu}=0$ are excluded from consideration
since group classification of all second-order linear PDEs in two dimensions was performed by Lie (see~\cite{vaneeva:Lie1881}).

Equations from this class are used to model various phenomena such as microwave heating, problems in population genetics, etc. (see, e.g.,~\cite{vaneeva:Bradshaw-Hajek2007} and references therein).
Theorems on existence and uniqueness of bounded solutions for equations of the more general form
$u_t=u_{xx}+F(t,x,u)$ were proven in~\cite{vaneeva:Kolmogorov}.

Lie symmetries of certain subclasses of~\eqref{eqRDk} are known. The group classification of
constant coefficient equations from class~\eqref{eqRDk} were studied by Dorodnitsyn~\cite{vaneeva:Dorodnitsyn1979} (the results are adduced in handbook~\cite{vaneeva:Ibragimov1994V1}). Class~\eqref{eqRDk} includes the generalized Huxley equations
\[u_t=u_{xx}+h(x)u^2(1-u),\]
whose Lie symmetries were studied in~\cite{vaneeva:Bradshaw-Hajek2004,vaneeva:Ivanova2008}.
There exists also certain intersection with the results on group classification of the classes
\begin{gather}
u_t=u_{xx}+H(x)u^m+F(x)u,\quad m\neq0,1,\,H\neq0,\label{eq_power}\\
u_t=u_{xx}+H(x)u^2+G(x), \quad H\neq0,\label{eq_quadratic}
\end{gather}
which were obtained in~\cite{vaneeva:VPS2009}. Note that the group classifications for the general class of (1+1)-dimensional second-order quasilinear evolution equations \[u_t=F(t,x,u,u_x)u_{xx}+G(t,x,u,u_x), \quad F\neq0,\] was
carried out in~\cite{vaneeva:Basarab-Horwath&Lahno&Zhdanov2001}. Nevertheless those results obtained up to a very wide equivalence group seem to be
inconvenient to derive group classification for class~\eqref{eqRDk}.

The structure of this paper is as follows. In the next section we investigate equivalence transformations in class~\eqref{eqRDk}. Lie symmetries are classified  in Section~3.

\section{Equivalence transformations}
To solve a group classification problem for a class of differential equations
it is important to describe point transformations that
preserve the differential structure of the class and transform only its arbitrary elements.
Such transformations are called \emph{equivalence transformations}
and form a group~\cite{vaneeva:Ovsiannikov1982}.
By Ovsiannikov, the equivalence group consists of the nondegenerate point transformations
of the independent and dependent variables and of the arbitrary elements of the class,
where transformations for independent and dependent variables do not involve arbitrary
elements, i.e., they are projectible on the space of independent and dependent variables.
If this restriction is neglected, then equivalence group is called a generalized equivalence group~\cite{vaneeva:Meleshko1994}.
If new arbitrary elements appear to depend on old ones in some nonpoint (possibly, nonlocal) way,
then the corresponding equivalence group is called extended~\cite{vaneeva:mogran}.
Admissible transformation is a triple consisting of two fixed equations from a class and  a point transformation that links these equations.
The set of admissible  transformations of a class naturally possesses
the groupoid structure with respect to the composition of transformations,
and hence it is called the \emph{equivalence groupoid} of the class~\cite{vaneeva:Popovych&Bihlo2012}.
We look for admissible transformations for class~\eqref{eqRDk}
using the direct method~\cite{vaneeva:Kingston&Sophocleous1998,vaneeva:VPS2012}.

Consider a pair of equations from the class under consideration,
i.e., equation~\eqref{eqRDk} and the equation
\begin{equation}\label{eqRDktilde}
{\tilde u}_{\tilde t}={\tilde u}_{\tilde x\tilde x}
+\tilde h(\tilde x)\tilde B(\tilde u),
\end{equation}
and assume that they are connected via a point
transformation~$\mathcal T$  of the general form
\begin{equation}\label{adm_tr}
\tilde t=T(t), \quad \tilde x=X(t,x), \quad \tilde u=U(t,x,u),
\end{equation}
where $T_tX_xU_u\ne0$. We can restrict ourselves by the transformations of this form
instead of general transformations $\tilde t=\hat T(t,x,u), $ $\tilde x=\hat X(t,x,u),$ $\tilde u=\hat U(t,x,u).$
It is due to the fact that class~\eqref{eqRDk} is a subclass of more general class of (1+1)-dimensional quasi-linear evolution equations
$u_t=F(t,x,u)u_{xx}+G(t,x,u,u_x)$ with $F\ne 0$, for which admissible transformations are proven to be of the form~\eqref{adm_tr}~\cite{vaneeva:Popovych&Ivanova2004NVCDCEs}.

We have to derive the determining equations for the functions~$T$, $X$ and $U$ and then to solve them.
Simultaneously we have to find a connection between arbitrary elements
of equations~\eqref{eqRDk} and~\eqref{eqRDktilde}.
Substituting the expressions for the new (tilded) variables
into~\eqref{eqRDktilde}, we obtain an equation in the old (untilded) variables.
It should be an identity on the manifold~$\mathcal L$ determined by~\eqref{eqRDk}
in the second-order jet space~$J^2$ with the independent variables $(t,x)$ and the dependent variable~$u$.
To involve the constraint between variables of~$J^2$ on the manifold~$\mathcal L$,
we substitute the expression of~$u_t$ implied by equation~\eqref{eqRDk}.
The splitting of this identity with respect to the derivatives $u_{xx}$ and $u_x$
implies the determining equations for the functions~$T$, $X$ and $U$:
\begin{gather}
U_{uu}=0,\quad{X_x}^2=T_t,\quad 2\frac{U_{xu}}{U_u}=-\frac{X_t
 X_x}{T_t}+\frac{X_{xx}}{X_{x}},\label{1DetEqForClassificationOfAdmTrans}
\\[1ex]
T_t\tilde h\tilde B- U_u hB=U_t -U_{xx}-\frac{X_t}{X_x}U_x.\label{2DetEqForClassificationOfAdmTrans}
\end{gather}
Solving at first equations~\eqref{1DetEqForClassificationOfAdmTrans} we get that  $T_t>0$ and
\begin{gather*}
X=\varepsilon\sqrt{T_t}\, x+\sigma(t),\quad U=U^1(t,x)u+U^0(t,x),\\
U^1=\zeta(t)\exp\left(-\frac18\frac{T_{tt}}{T_t}x^2-\frac\varepsilon2\frac{\sigma_t}{\sqrt{T_t}}x\right),
\end{gather*}
where  $U^0$, $\sigma$ and  $\zeta$ are arbitrary smooth functions of their variables; $ \varepsilon=\pm1$. Then~\eqref{2DetEqForClassificationOfAdmTrans}
can be written as
\begin{gather}\label{eq:hBconnection}
{T_t}\tilde h\tilde B- U^1 h B=
\sum^1_{i=0}\left(U^i_t-U^i_{xx}-\dfrac12\dfrac{T_{tt}}{T_t}U^i_xx-
\varepsilon\dfrac{\sigma_t}{\sqrt{T_t}}U^i_x\right)u^i,
\end{gather}
where $u^1=u,$ $u^0=1.$
Investigating~\eqref{eq:hBconnection}  for varying arbitrary elements $h$ and $B$, we derive at first the usual equivalence group of class~\eqref{eqRDk}.
\begin{thm}\label{equiveqRDk}
The usual equivalence group~$G^{\sim}$ of
class~\eqref{eqRDk} consists of the transformations
\[
\begin{array}{l}
\tilde t=\delta_1^2 t+\delta_2,\quad \tilde x=\delta_1x+\delta_3, \quad
\tilde u=\delta_4u+\delta_5, \quad
\tilde h=\dfrac{\delta_4}{\delta_1^2\delta_0} h, \quad
\tilde B=\delta_0B,
\end{array}
\]
where  $\delta_j,$ $j=0,\dots,5,$  are arbitrary constants with
$\delta_0\delta_1\delta_4\not=0$.
\end{thm}
It appears that there exist point transformations between equations from~\eqref{eqRDk}
 which do not belong to~$G^{\sim}$ and
form a
conditional equivalence group. Moreover, this group is not usual but a generalized extended one.

\begin{thm}\label{equiveqRDk2}
The generalized extended equivalence group~$\hat G^{\sim}_{\exp}$ of the
subclass
\begin{equation}\label{eq_expr}
u_t=u_{xx}+h(x)(e^{nu}+r)
\end{equation}
of class~\eqref{eqRDk} is formed by the transformations
\[
\begin{array}{l}
\tilde t=\delta_1^2 t+\delta_2,\quad \tilde x=\delta_1x+\delta_3, \quad
\tilde u=\delta_4u+\varphi(x), \\[1ex]
\tilde h=\dfrac{\delta_4}{\delta_1^2}e^{-\frac{n}{\delta_4}\varphi} h, \quad\tilde n=\dfrac{n}{\delta_4},\quad
\tilde r=e^{\frac{n}{\delta_4}\varphi}\left(r-\dfrac{\varphi_{xx}}{\delta_4h}\right),
\end{array}
\]
where  $r$ and $\delta_j,$ $j=1,\dots,4,$  are arbitrary constants with
$\delta_1\delta_4\not=0$. The transformation component for $r$ can be interpreted as the constraint for $\varphi$, \[\varphi_{xx}=\delta_4h(r-\tilde re^{-\frac{n}{\delta_4}\varphi}).\]
\end{thm}

Theorem~\ref{equiveqRDk2} implies that class~\eqref{eq_expr} reduces to the class
\begin{equation}\label{eq_exp}
\tilde u_t=\tilde u_{xx}+\tilde h(x)e^{n\tilde u}
\end{equation}
 by the transformation
\[
\tilde t=t, \quad\tilde x=x,\quad\tilde u=u+\varphi(x),
\]
where $\tilde h(\tilde x)=e^{-\varphi(x)} h(x)$ and $\varphi_{xx}=rh(x).$
Class~\eqref{eq_expr} is normalized. Therefore, the equivalence group of class~\eqref{eq_expr} with $r=0$ can be found
setting $\tilde r=r=0$ in transformations from the group~$\hat G^{\sim}_{\exp}$.
\begin{cor}\label{equiveqRDk3}
The usual equivalence group~$G^{\sim}_{\rm exp}$ of the class
\[
u_t=u_{xx}+h(x)e^{nu}
\]
 consists of the transformations
\[
\begin{array}{l}
\tilde t=\delta_1^2 t+\delta_2,\quad \tilde x=\delta_1x+\delta_3, \quad
\tilde u=\delta_4u+\delta_5x+\delta_6, \\[1ex]
\tilde h=\dfrac{\delta_4}{\delta_1^2}e^{-\frac{n}{\delta_4}(\delta_5x+\delta_6)} h, \quad\tilde n=\dfrac{n}{\delta_4},
\end{array}
\]
where  $\delta_j,$ $j=1,\dots,6,$  are arbitrary constants with
$\delta_1\delta_4\not=0$.
\end{cor}

In the course of the study of Lie symmetries we will use the derived equivalence transformations for the simplification of calculations and for
presenting the final results in a concise form.

\section{Lie  symmetries}
We study Lie symmetries of equations from class~\eqref{eqRDk} using the classical approach~\cite{vaneeva:Ovsiannikov1982} in combination with  the method of furcate split~\cite{vaneeva:Popovych&Ivanova2004NVCDCEs}.
We search for vector fields of the form
\begin{equation*}
Q=\tau(t,x,u)\partial_t+\xi(t,x,u)\partial_x+\eta(t,x,u)\partial_u
\end{equation*}
that generate one-parameter Lie symmetry groups of a fixed equation~$\mathcal L$ from class~\eqref{eqRDk}.
These vector fields form the maximal Lie invariance algebra $A^{\max}=A^{\max}(\mathcal L)$ of the equation~$\mathcal L$.
Any Lie symmetry generator~$Q$ satisfies the infinitesimal invariance criterion, i.e.,
the action of the second prolongation~$Q^{(2)}$ of~$Q$ on the equation~$\mathcal L$
results in the condition  identically satisfied for all solutions of~$\mathcal L$.
Namely, we require
\begin{equation}\label{c1}
Q^{(2)}\big(u_t-u_{xx}-h(x)B(u)\big)\Big|_{\mathcal L}=0.
\end{equation}

After elimination of $u_t$ by means of~\eqref{eqRDk},
equation~\eqref{c1} can be regarded as a polynomial in the variables $u_x$, $u_{xx}$ and $u_{tx}$.
The coefficients of different powers of these variables should be zeros.
This results in the determining equations for the coefficients $\tau$, $\xi$ and $\eta$.
Solving these equations implies that $\tau =\tau(t)$ and $\xi =\xi(t,x)$, which
agrees with the general results on point transformations between evolution equations~\cite{vaneeva:Kingston&Sophocleous1998}.
The remaining determining equations have the form
\begin{gather}
2\xi_x=\tau_t,\qquad
\eta_{uu}=0,\qquad
2\eta_{xu}=\xi_{xx}-\xi_t,\label{eq_det1}\\
\eta h B_u=\left(-\xi{h_x}+\left(\eta_u-\tau_t\right)h\right) B+\eta_t-\eta_{xx}.\label{eq_det2}
\end{gather}
Integrating equations~\eqref{eq_det1} we get the following expressions for $\xi$ and $\eta$:
\begin{equation*}
\xi=\tfrac 12 \tau_t\, x+\sigma(t), \quad
\eta=\left(-\tfrac 18 \tau_{tt} x^2-\tfrac 12\sigma_t x+\zeta(t)\right)u+\eta^0(t,x),
\end{equation*}
where $\sigma$, $\zeta$ and $\eta^0$ are arbitrary smooth functions of their variables. Then equation~\eqref{eq_det2} becomes
\begin{gather}\arraycolsep=0ex\label{eq_classifying}
\begin{array}{l}
 \left(\bigl( \tfrac18\tau_{tt}x^2+\tfrac12\sigma_tx-\zeta\bigr)u-\eta^0\right)hB_u= \left(\left(\tfrac18\tau_{tt}x^2+\tfrac12\sigma_tx-\zeta+
 \tau_t\right)h+{}\right.\\[1ex]+\left.\left(\tfrac12\tau_tx+\sigma\right){h_x}\right)B+\left (\tfrac18\tau_{ttt}x^2+\tfrac12\sigma_{tt}x-\zeta_t-\tfrac14\tau_{tt}\right) u-\eta^0_t+
 \eta^0_{xx}.
 \end{array}
\end{gather}
It is called a classifying equation and should be solved simultaneously with respect to remaining uncertainties in the coefficients of infinitesimal generator $Q,$ i.e., the functions $\tau$, $\sigma$, $\zeta$ and $\eta^0$, and arbitrary elements of the class, namely, $h$ and $B$.

In order to find the common part of Lie symmetries for all equations from class~\eqref{eqRDk},
we split with respect to the arbitrary elements in equation~\eqref{eq_classifying}.
This results in $\tau_t=\xi=\eta=0$.
\begin{prop}
The intersection of the maximal Lie invariance algebras of equations
from class~\eqref{eqRDk} (called kernel algebra) is the one-dimensional algebra
$A^\cap=\langle\partial_t\rangle$.
\end{prop}

The next step is to classify  possible extensions of~$A^\cap$ using the method of furcate
split~\cite{vaneeva:VPS2012,vaneeva:Popovych&Ivanova2004NVCDCEs}.
For any operator $Q$ from $A^{\max}$ the substitution of its
coefficients into~\eqref{eq_classifying}  gives some equation on $B$
of the general form
\begin{gather}\label{eq_lie_sym_fc}
(a u+b)B_u=p B +qu+r,
\end{gather}
where $a,$ $b,$ $p,$ $q$ and $r$ are constants which are defined up to nonzero multiplier.
The set~$\mathcal V$ of values of the coefficient tuple $(a,b,p,q,r)$
obtained by varying of an operator from $A^{\max}$ is a linear space.
Note that $(a,b)\neq(0,0)$ and
the dimension $k=k(A^{\max})$ of the space~$\mathcal V$ is not greater than 2.
Otherwise the corresponding equations imply that either $B$ is linear in $u$ or the system is incompatible.
The value of $k$ is an invariant of the transformations from $G^\sim$.
Therefore, there exist three $G^\sim$-inequivalent cases
for the value of $k$: $k = 0,$ $k = 1$ and $k = 2$.
We consider these possibilities separately.

\noindent{\bf I.}
The condition $k=0$ means that \eqref{eq_classifying} is not an equation with respect to $B$ but an identity.
Therefore, $B$ is not constrained and $\tau_{tt}=\sigma_t=\zeta=\eta^0=0$.
We obtain that
$\tau=c_1t+c_2$, $\sigma=c_3$ and the classifying equation on $h$
\begin{equation}\label{eq_classifying2}
\left(\tfrac12c_1x+c_3\right)h_x+c_1h=0,
\end{equation}
where $c_1,$ $c_2$, and $c_3$ are arbitrary constants. If $h$ is arbitrary we get the kernel algebra~$A^\cap$ presented by Case~0 of Table~1.
It follows from~\eqref{eq_classifying2} that the extensions are possible in two cases: either
 $h=\delta (x+\beta)^{-2}$ or  $h=\delta$, where $\beta$ and $\delta$ are arbitrary constants, $\delta\neq0$. Up to $G^\sim$-equivalence $\beta$
 can be set to zero value and $\delta$ to $\pm1$ depending on its sign. These two cases are presented by Cases~1 and 2 of Table 1.

\noindent{\bf II.}
If $k=1$ then we have, up to a nonzero multiplier,
exactly one equation of the form~\eqref{eq_lie_sym_fc} with respect to the function $B$.
Integration of this equation up to $G^\sim$-equivalence gives three cases
\begin{equation*}
1.\,\,B=u^m+\beta_1u+\beta_2,\quad 2.\,\,B=e^u+\beta_1u+\beta_2,\quad 3.\,\, B=u\ln u+\beta_2,
\end{equation*}
where $\beta_1, \beta_2$ and $m$ are arbitrary constants, $m\neq0,1.$ The next step is to substitute each of the three derived forms of $B$ to~\eqref{eq_classifying} and subsequently to split it with respect to linearly independent functions of the variable $u$. Consider these cases separately.

\noindent{\bf II.1.} Let $B=u^m+\beta_1u+\beta_2$. If  $m\neq2$ then the splitting equation~\eqref{eq_classifying} leads to the condition $\eta^0=0$
and the three classifying equations on the function $h$
\begin{gather*}\label{eq_classifying2a}
\left(\tfrac12\tau_tx+\sigma\right)h_x+\left((1-n)W+
 \tau_t\right)h=0,\quad \beta_2\left(\left(\tfrac12\tau_tx+\sigma\right)h_x+\left(W+
 \tau_t\right)h\right)=0,\\
(1-\beta_1)\left(\tfrac12\tau_tx+\sigma\right)h_x+\left(W+(1-
 \beta_1)\tau_t\right)h-W_t+\tfrac14\tau_{tt}=0,
\end{gather*}
where
$
W=\tfrac18\tau_{tt}x^2+\tfrac12\sigma_tx-\zeta,$ $W_t=\frac{\partial W}{\partial t}.
$
The studying of compatibility of this system shows that if $\beta_2\neq0$ then $\tau=c_1t+c_2$, $\sigma=c_3$, $\zeta=0$ and the classifying equation on $h$
is of the form~\eqref{eq_classifying2}. Therefore, this is subcase of the case $k=0$ considered above.

 If $m\neq 2$ and $\beta_2=0$ then class~\eqref{eqRDk} takes the form~\eqref{eq_power} with $H(x)=h(x)$ and $F(x)=\beta_1h(x)$. Using the results of~\cite{vaneeva:VPS2009} we derive that extension of $A^\cap$ is possible  if $\beta_1=0$ and the function $h$ take one of the forms presented
by  Cases 3, 4 and 5 of Table 1. If $\beta_1\neq0$ then the cases of extension of $A^\cap$ are subcases of Cases~1 and 2 of Table~1.

 If $B=u^2+\beta_1u+\beta_2$, then $\beta_1$ can always be set to zero by the translation $u\mapsto u-\beta_1/2$ and then class~\eqref{eqRDk} coincides with~\eqref{eq_quadratic}, where $H(x)=h(x)$ and $G(x)=\beta_2h(x)$. Using the results derived in~\cite{vaneeva:VPS2009} we obtain that the cases of Lie symmetry extension are either presented by Cases 1 and 2 of Table 1
if $\beta_2\neq0$ or by Cases 3--5 of Table 1
 if $\beta_2=0$.

\noindent{\bf II.2.} If $B=e^u+\beta_1u+\beta_2$, then the classifying equations imply
$\tau_{tt}=\sigma_t=\zeta=0$.
We obtain that
$\tau=c_1t+c_2$, $\sigma=c_3$ and the system of equations on $h$
\begin{gather*}\label{eq_classifying3}
\left(\left(\tfrac12c_1x+c_3\right)h_x+c_1h\right)\beta_1=0,\quad \eta^0h+\left(\tfrac12c_1x+c_3\right)h_x+c_1h=0,\\
\left(\left(\tfrac12c_1x+c_3\right)h_x+c_1h\right)\beta_2+\eta^0h\,\beta_1+\eta^0_{xx}-\eta^0_t=0.
\end{gather*}
If $\beta^1\neq0$, then $\eta^0=0$ and the classifying condition on $h$ is exactly equation~\eqref{eq_classifying2}. This case can be included to the general case with arbitrary $B$. If $\beta^1=0$, then $\beta_2$ can be set to zero using the transformation from the conditional equivalence group~$\hat G^\sim_{\rm exp}$. Then $\eta^0=c_4x+c_5,$ where $c_4$ and $c_5$ are arbitrary constants and the remaining classifying equation on $h$ takes the form
\begin{equation}\label{eq_classifying5}
\left(\tfrac12c_1x+c_3\right)h_x+(c_4x+c_5+c_1)h=0.
\end{equation}
Combined with the multiplication by a nonzero constant,
each transformation from the equivalence group~$G^{\sim}_{\rm exp}$ is extended to the coefficient tuple
of the above equation in the following way
\begin{gather*}
\tilde c_1=\kappa\, c_1,\quad \tilde c_3=\kappa\left(c_3\delta_1-\tfrac12 c_1\delta_3\right),\quad \tilde c_4=\frac{\kappa}{\delta_1}\left(c_4+\tfrac12 c_1\delta_5\right),\\ \tilde c_5=\frac{\kappa}{\delta_1}\left(c_5\delta_1+c_3\delta_1\delta_5-c_4\delta_3-\tfrac1 2{c_1}\delta_3\delta_5\right).
\end{gather*}
Here $\kappa$ is an arbitrary nonzero constant. Using these transformations we derive the following statement.
\begin{lem}\label{LemmaOntransOfCoeffsOfClassifyingSystem}
Up to $G^{\sim}_{\rm exp}$-equivalence
the parameter tuple~$(c_1,c_3,c_4,c_5)$ can be assumed to belong to the set
$
\{(1,0,0,\bar c_5),\ (0,1,\pm 2,0),\ (0,1,0,0),\},
$
where $\bar c_5$ is an arbitrary constant.
\end{lem}
Integration of~\eqref{eq_classifying5} up to $G^{\sim}_{\rm exp}$-equivalence leads to the forms of $h$ presented in Cases 6--8 of Table~1.
In  Case 6 we use the notation $s=-2(\bar c_5+1).$

\noindent{\bf II.3.} If $B=u\ln u+\beta_2$, then $\eta^0=0$ and the remaining classifying conditions are
\begin{gather*}\label{eq_classifying4}
\left(\tfrac12\tau_tx+\sigma\right)h_x+\tau_th=0,\quad
\left(\tfrac18\tau_{tt}x^2+\tfrac12\sigma_tx-\zeta\right)h=\tfrac18\tau_{ttt}x^2+\tfrac12\sigma_{tt}x-\zeta_t-\tfrac14\tau_{tt},\\
\beta_2\left(\left(\tfrac12\tau_tx+\sigma\right)h_x+\left(\tfrac18\tau_{tt}x^2+\tfrac12\sigma_tx-\zeta+
 \tau_t\right)h\right)=0.
\end{gather*}
Investigation of this system  implies that if $\beta_2\neq0$ or
 $\beta_2=0$ but $h_x\neq0$, then $\tau_{tt}=\sigma_t=\zeta=0$. Therefore
$\tau=c_1t+c_2$, $\sigma=c_3$ and the classifying equation on $h$ is~\eqref{eq_classifying2}. I.e. we get nothing but subcases of Cases 1 and 2 of Table 1.
But if $\beta_2=0$ and $h=\delta={\rm const}$, then the classifying equations leads to the conditions $\tau_t=\sigma_{tt}-\delta\sigma_t=\zeta_t-\delta\zeta=0$. Therefore
$\tau=c_1$, $\xi=c_2+c_3e^{\delta t},$ $\eta=\left(-\tfrac{\delta}2c_3x+c_4\right)e^{\delta t}u$ and we have extension of $A^\cap$ on three Lie symmetry operators (Case 9 of Table~1).
\begin{table}[t]
\begin{center}
\renewcommand{\arraystretch}{1.2}
\textbf{Table~1.}
The group classification of the class $u_t=u_{xx}+h(x)B(u)$,  $hB_{uu}\neq0$.
\\[2ex]
\begin{tabular}{|c|c|c|l|}
\hline
no.&$B(u)$&$h(x)$&\hfil Basis of $A^{\max}$ \\
\hline
0&$\forall$&$\forall$&$\partial_t$\\
\hline
1&
$\forall$&$\delta x^{-2}$&$\partial_t,\,2t\partial_t+x\partial_x$\\
\hline
2&
$\forall$&$\delta $&$\partial_t,\,\partial_x$\\
\hline
3&
$u^{m}$&$\delta x^s$&$\partial_t,\,2(m-1)t\partial_t+(m-1)x\partial_x-(s+2)u\partial_u$\\
\hline
4&
$u^{m}$&$\delta e^x$&$\partial_t,\,(1-m)\partial_x+u\partial_u$\\
\hline
5&
$u^{m}$&$\delta $&$\partial_t,\,\partial_x,\,2(m-1)t\partial_t+(m-1)x\partial_x-2u\partial_u$\\
\hline
6&
$e^{u}$&$\delta x^s$&$\partial_t,\,2t\partial_t+x\partial_x-(s+2)\partial_u$\\
\hline
7&
$e^{u}$&$\delta e^{\pm x^2}$&$\partial_t,\,\partial_x\mp 2x\partial_u$\\
\hline
8&
$e^{u}$&$\delta $&$\partial_t,\,\partial_x,\,2t\partial_t+x\partial_x-2\partial_u$\\
\hline
9&
$u\ln u$&$\delta $&$\partial_t,\,\partial_x,\, e^{\delta t}u\partial_u,\, e^{\delta t}(\partial_x-\frac{\delta}2xu\partial_u)$\\
\hline
\end{tabular}
\\[2ex]
\parbox{120mm}{Here $\delta,$ $m$ and $s$ are arbitrary constants, $m\neq0,1,$ $s\neq 0,$ $\delta=\pm1\bmod G^\sim$}
\end{center}
\end{table}

\noindent{\bf III.} Let $k=2$. We choose a basis $\{(a_i,b_i,p_i,q_i,r_i),\ i=1,2\}$
of the space~$\mathcal V$ of tuples $(a,b,p,q,r)$ associated with~$A^{\max}$. The system on $B$ is of the form
\begin{gather*}
(a_1 u+b_1)B_u=p_1 B +q_1u+r_1,\\
(a_2 u+b_2)B_u=p_2 B +q_2u+r_2.
\end{gather*}
The determinant of the matrix  $M=\begin{pmatrix}
a_1 &b_1\\
a_2  &b_2
\end{pmatrix}
$ is nonzero, since otherwise $B$ is linear in $u$. Therefore the system can be rewritten in the form
\begin{gather*}
uB_u=p'_1 B +q'_1u+r'_1,\\
B_u=p'_2 B +q'_2u+r'_2,
\end{gather*}
where $\begin{pmatrix}
p'_1\\
p'_2
\end{pmatrix}=M^{-1}\begin{pmatrix}
p_1\\
p_2
\end{pmatrix},$ $\begin{pmatrix}
q'_1\\
q'_2
\end{pmatrix}=M^{-1}\begin{pmatrix}
q_1\\
q_2
\end{pmatrix},$ $\begin{pmatrix}
r'_1\\
r'_2
\end{pmatrix}=M^{-1}\begin{pmatrix}
r_1\\
r_2
\end{pmatrix};$ $M^{-1}$ is the inverse matrix of the matrix $M.$
Studying the latter system we get that
it is compatible only if
$B$ is either linear or quadratic. The case $B$ is quadratic in $u$ is considered already in the course of the study of the case $B=u^m+\beta_1u+\beta_2.$
So, the case $k=2$ does not lead to any new case of Lie symmetry extension for equations~\eqref{eqRDk} with $B_{uu}\neq0$.

The group classification problem for class~\eqref{eqRDk} is solved exhaustively. The results are summarized in Table~1.
It is important to note that group classification of subclass~\eqref{eq_expr} is carried out up to the $\hat G^\sim_{\rm exp}$-equivalence, whereas all other cases are classified up to the usual $G^\sim$-equivalence.

\section{Conclusion}
In this paper we solve the group classification problem for the class of (1+1)-dimensional quasilinear diffusion equations with a variable coefficient nonlinear source~\eqref{eqRDk} which arise as mathematical models in problems of mathematical biology and other applied areas~\cite{vaneeva:Bradshaw-Hajek2007}. The results of group classification can be applied for searching closed form solutions  via the classical reduction method. The knowledge of Lie symmetries is also necessary  for finding nonclassical symmetries (called also $Q$-conditional symmetries or reduction operators) of equations~\eqref{eqRDk}. This will be a subject of a forthcoming paper.

\subsection*{Acknowledgement} The authors are grateful to  Roman Popovych and Iryna Yehorchenko for useful discussions and constructive suggestions.
OV thanks the Organizing Committee of the XVII Geometrical Seminar for the hospitality and support.

\bibliographystyle{amsplain}

\end{document}